# Immersive Stories for Health Information: Design Considerations from Binge Drinking in VR

Douglas Zytko[1], Zexin Ma[2], Jacob Gleason[1], Nathaniel Lundquist[1], and Medina Taylor[2]

[1] Department of Computer Science and Engineering, Oakland University, Rochester, MI, USA
`zytko@oakland.edu`
[2] Department of Communication, Journalism, and Public Relations, Oakland University, Rochester, MI, USA
`zexinma@oakland.edu`

**Abstract.** Immersive stories for health are 360° videos that intend to alter viewer perceptions about behaviors detrimental to health. They have potential to inform public health at scale, however, immersive story design is still in early stages and largely devoid of best practices. This paper presents a focus group study with 147 viewers of an immersive story about binge drinking experienced through VR headsets and mobile phones. The objective of the study is to identify aspects of immersive story design that influence attitudes towards the health issue exhibited, and to understand how health information is consumed in immersive stories. Findings emphasize the need for an immersive story to provide reasoning behind a character's engagement in the focal health behavior, to show the main character clearly engaging in the behavior, and to enable viewers to experience escalating symptoms of the behavior before the penultimate health consequence. Findings also show how the design of supporting characters can inadvertently distract viewers and lead them to justify the detrimental behavior being exhibited. The paper concludes with design considerations for enabling immersive stories to better inform public perception of health issues.

## 1 Introduction

*"Nothing like seeing a group of doctors smoking outside to let you know that information alone doesn't persuade."* – Ramit Sethi

Urban legends told over campfire, figures etched in old stone, an article in a yellowing newspaper, a comic book, a novel, a radio program, a movie, a Youtube video. Story has been a quintessentially human way of sharing information for centuries, and the most powerful stories do more than just inform - they alter, persuade, and culminate in action.

Technology has continually expanded the ways in which stories are told and impact our behavior. Immersive stories [1] are a particularly novel form of storytelling; they utilize 360° video to share narrative through virtual reality [2] headsets and mobile devices. Through immersive stories a viewer typically experiences a narrative in first person through the eyes of a character and can freely explore their virtual surroundings by manipulating their character's gaze. This immersion can instill a sense of presence (or "being there") and the perception that one is personally experiencing a story's events [3, 4].

Immersive stories are growing in popularity for news consumption [5] and for informing the public about health issues [6]. The potential for viewers to viscerally experience the implications of a public health issue can be a powerful way to not only shape understanding, but to convince the public to adopt healthier behavioral choices [7]. This capacity for attitudinal and behavioral change is of paramount and timely importance in light of health pandemics that necessitate mass adoption of specific health behaviors to keep the public safe.

Despite the potential of immersive stories to inform and improve public health at scale, knowledge of immersive story design is still in fledgling stages. A majority of immersive story research has explored differences in viewing device [8-14] (e.g., VR headsets, phones, and computer monitors), leaving immersive story content and design as a figurative "black box" that is relatively understudied. Some literature has identified and provided opinions on the design of discrete story elements—namely character perspective choice [15-19] and direction of viewer attention [10, 12, 14, 20-23]. However, actual viewers of immersive stories have largely been absent from design research. Their involvement is vital not just for validating the impact of pre-identified design elements [18, 21, 22], but for *identifying* immersive story elements impactful to their perceptions and articulating how design choices impact them. This gap is especially pronounced for immersive stories in the public health context: what elements of immersive story design are impactful to viewers' attitudes towards a respective public health issue, and how do they impact viewers?

To address this gap, we conducted a focus group study with viewers of an immersive story about binge drinking—a widespread and severe public health issue [24, 25]—to identify and explore elements of immersive story design critical to viewer attitudes about the respective health issue. In the next section we review prior research on immersive stories and their design, which informs our research questions and method. We then present the focus group study, and conclude by discussing considerations and directions for immersive story design based on the study's findings.

## 2   Background

Immersive stories are video-based stories that offer a 360° panoramic view and spatialized audio [26, 27]. They are typically viewed on VR headsets and mobile devices, as facilitated with special viewing options on Youtube and other media outlets, and have been increasingly used in journalism, marketing, and non-profit sectors [1]. For example, mainstream media like *The New York Times*, *the BBC*, and *USA Today* have used immersive stories to offer experiential insight into news events [1, 5]. Organizations like the *United Nations* have used immersive stories to promote social good [28].

Immersive stories are a promising tool for improving health-related knowledge and facilitating behavior change [7] because they allow viewers to vividly experience health implications and symptoms of illness without suffering from any serious harm. An example is *A Walk Through Dementia*, which tells stories of people living with different forms of dementia by putting viewers in the shoes of those battling the disease [29].

As a relatively new storytelling experience, scientific research into immersive stories is still in its infancy. The majority of research has focused on the role of modality by comparing the viewing experience across head-mounted displays (HMDs), mobile devices, and flat computer screens [5, 8-14, 30-32]. These studies reinforce the potential of immersive stories, but with mixed results as to how choice of viewing device affects viewer experience—if at all [11].

There is relatively limited understanding of best practices for designing immersive stories, or what Banos and colleagues call "elements of the [virtual] environment and the content itself" [13]. Furthermore, immersive story research has focused on varying outcomes or goals of the viewing experience—including presence [4, 5, 8, 10, 11, 13, 14], enjoyment [5], empathy [11, 14, 17, 30, 33], and emotional response [8, 9, 13]—making it difficult to synthesize design insight for health contexts. Attitudinal and behavioral change (the goals of health-related stories) have been particularly rare focal outcomes of immersive story research [8].

Early research into immersive story design has focused predominantly on character perspective and directing viewer attention. Directing viewer attention has been recognized as a primary challenge to immersive story design due to viewers having the freedom to manipulate their gaze, therefore making it possible to miss key plot elements [10, 12, 14, 20-22]. Research has posed and tested various audio-based, visual-based, and mixed audio/visual cues for directing viewer attention [14, 21-23].

Regarding character perspective, research has identified three options available to designers. One is a *third person/non-character observer* who plays no role in the story [17-19]. The other two are first-person character perspectives (the viewer experiences the story through the eyes of a character) including the *main character* of the story who engages in plot-driving actions and a *supporting character* who observes plot-driving occurrences [15, 17]. There is some empirical evidence arguing against the non-character observer role or for a first-person perspective more broadly [11, 34-36]. However, there is a lack of empirical evidence into the main character and supporting character choices—speculations on the advantages of each character perspective can be found in [15, 17].

A rare example of immersive story design research beyond character perspective and user attention direction involved the identification of story elements broadly conducive to immersion [16], although these are based on the researchers' perspective rather than insight from viewers. The only research into immersive story design pertinent to viewer attitudes around a health issue, to the authors' knowledge, studied the impact of broadly conceptualized "emotional content" [8].

Ultimately, there persists a need for empirical research that identifies elements of immersive story design that viewers consider impactful to their experience, especially their attitudes towards exhibited public health issues. Our study pursued the following research questions:

1. What elements of immersive story design do viewers consider influential to their attitudes towards the health information being conveyed?
2. How do these elements impact viewers' perceptions of the health information?
3. In what ways could immersive story design for health be improved so that health information impacts viewers' attitudes as intended?

## 3  Method

To explore the research questions, we conducted a focus group study with 147 participants about an immersive story depicting binge drinking and its negative health consequences. Participants were graduate and undergraduate students (mean age = 22.14, SD = 4.05) recruited from a medium-sized public university in the Midwest of the United States, and were compensated with extra course credit. Participants identified as White (77.6%), Black (10.9%), Asian (6.8%), Hispanic (2.7%), and other (2.0%). Regarding gender, 50.3% identified as male and 49.7% as female.

### 3.1  The Immersive Story

Binge drinking was chosen as the focal health issue based on its severity and prevalence [24], particularly amongst our participant demographic of young adults [25]. Binge drinking is a pattern of drinking alcoholic beverages that brings a person's blood alcohol concentration (BAC) to 0.08 g/dL; typically 4+ drinks for women and 5+ for men [37].

The immersive story about binge drinking used in the study [38, 39] was selected because it represents common immersive story design choices: a first-person point of view, the ability to select a character perspective, gaze manipulation around a 360° field of vision, and a duration that is 10 minutes or less [15] (this particular story is 6 minutes in duration).

The story centers on Greg, who engages in binge drinking at a house party hosted by his friend Stephanie because he is moving away for a new job. Greg drinks 11 alcoholic beverages over a 5-hour period (a mix of beer, wine, and shots), which is conveyed by intermittent cuts to a black screen that shows a number of alcoholic beverages being added to a table with a time stamp. Greg's intoxication is exhibited with blurred and slowed vision later in the story. Stephanie observes and interacts with Greg throughout the story, eventually helping him to a couch to fall asleep. In the morning Stephanie finds Greg unresponsive due to alcohol poisoning and calls an ambulance.

The video opens with text content about the definition of binge drinking, and its potential health consequences including alcohol poisoning. After the immersive story is experienced the video ends with text information about the severity of binge drinking.

There are two versions of the immersive story that vary based on character perspective, but otherwise have identical plots and information. In one version, the viewer experiences the story through the eyes of Greg, and in the other through the eyes of Stephanie.

### 3.2 Data Collection and Analysis

Focus group discussions were the second stage in a broader study about immersive stories for health; the first stage (an experiment on character perspective) necessitated that participants experience the binge drinking immersive story on Youtube either through a VR headset or smart phone, and either from Greg's perspective or Stephanie's perspective (rendering 4 different conditions that participants were evenly allocated to). On a smartphone the viewer moves the handheld device in any direction to manipulate their character's gaze; on the VR headset the viewer moves their head to manipulate gaze (See Fig. 1). All participants used headphones for the immersive story's audio. Focus groups were conducted on the university campus in two separate rooms immediately after participants viewed the immersive story and filled out a survey regarding the experience for the first stage of the study. One room was for participants who viewed the immersive story through a smart phone, and the other for VR headsets. The 147 participants were split across 15 focus group sessions, ranging from 4 to 20 participants each. Session lengths ranged from 15 to 25 minutes, as determined by level of disagreement or debate amongst participants.

Focus group discussions opened with participants describing the overall experience of viewing the immersive story, and then recollecting (and sometimes debating) the plot of the immersive story in their own words. Participants were prompted to discuss their attitudes and understanding of the story, the characters (Greg and Stephanie), and the health-related behavior exhibited (binge drinking).

All focus group discussions were voice recorded and transcribed. A team of five researchers held recurrent, synchronous sessions for inductive open coding of the transcripts [40], followed by organization of emergent themes using axial coding, and then finalization of the code mapping and hierarchy.

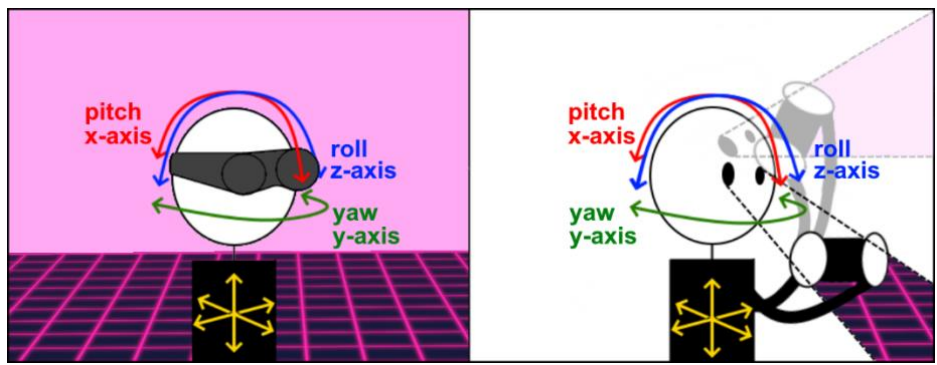

**Fig. 1.** Manipulating character gaze on a VR headset (left) versus a smartphone (right)

## 4      Findings

Participants' attitudes and perceptions towards the health information provided in the binge drinking immersive story were mixed. Some expressed a reluctance to engage in excessive drinking after watching. Others, however, were skeptical of the health impacts of binge drinking as depicted in the immersive story, and were either confused by the behavior or interpreted the behavior as justified.

Participants referenced two overarching aspects of immersive story design as influential to their attitudes and perceptions about binge drinking: 1) *health behavior design*, which comprises aspects of an immersive story that serve to convey and explain the focal health behavior; and 2) *supporting character design*, which comprises the actions of a character who observes the detrimental health behavior (but does not personally engage in it) and how that character's role is understood by viewers. The coding process did not reveal any themes unique to viewing the immersive story in a VR headset or mobile phone. There were some differences based on character perspective, which are noted throughout the findings.

### 4.1      Health Behavior Design

Axial coding partitioned health behavior design into three sub-concepts: the context of the focal behavior (where and why it happens), engagement in the behavior, and health consequences of the behavior.

**Context of the Behavior.** Most participants understood that Greg was drinking a large amount of alcohol, but they reported needing to know his reasons or *"backstory"* for excessive drinking in order to fully empathize with the character. This backstory would be distinct from contextual information that explains why characters would be drinking alcohol in general. Participants universally understood the going-away party context due to the house setting and party guests visible around the virtual environment. However they did not consider the virtual environment to offer much information to contextualize Greg's *excessive* drinking relative to other characters. In lieu of this information, participants invented their own backstories for Greg's drinking that usually involved a need to *"escape"* real world problems, such as family drama, fights with other partygoers, and sadness about moving away. These speculations were sometimes used to justify Greg's behavior and (as will be reported later) shift the blame for Greg's hospitalization to the party host, Stephanie.

*"A lot of people drink to escape, you don't know what he was going through in the video, it never exactly said why he was drinking so much. He was moving, he probably could've been sad about moving away." (Smartphone, Greg's perspective, female)*

*"And the other thing, one thing for me personally, that I just think would have added something to the level of empathy for the characters - if I knew their backstory, in the sense of like, why is Greg drinking so much? [Was it] a fight? Did he have a bad history with his family? Was there a reason for that?" (VR headset, Greg's perspective, male)*

Relatedly, some participants wanted to know the amount of alcohol that Greg typically drinks at parties, speculating that he may not have been aware of his limits. *"I* *would like to know if he got drunk more in the past, that same amount or was that his first time drinking a whole lot?" (VR headset, Greg's perspective, female)*

Several participants scanned the 360° environment for explanations of Greg's excessive drinking. Some assessed the party to consider how they would have acted in a similar situation. As one participant put it: *"The environment didn't feel joyous anyway, like it's a party but this seems like a really lame party. Like I would've been just drinking too" (Smartphone, Greg's perspective, male).* Critical perceptions of the party were influenced by the number of partygoers (approximately 10 people are seen throughout the house) and their activities (conversation and light dancing), which some participants considered incongruent with parties they had personally attended that involved heavy drinking.

A few of these participants expressed frustration with scanning the environment for contextual information. While there were various audible cues throughout the story prompting viewers to redirect their attention, some were hesitant to explore them because they feared missing vital plot information happening in their immediate field of vision. This hesitance to redirect attention was most pronounced in scenes when someone was directly interacting with the viewer's character, because it would normally be considered rude to redirect visual attention while in the middle of conversation.

*"And I think in terms of having a 360 experience, it's kind of strange because a lot of times when you're engaging with someone, you're engaging straight, you're not looking over here [referencing an area in the periphery]. Most people will be focused on what's going on in front of them, versus what's going on behind them." (VR headset, Stephanie's perspective, male)*

Some participants reported that they stopped reacting to audible cues around them because they found prior cues to be uninteresting or uninformative to the plot when they sought out their sources. *"…there was nothing really interesting to look at, like on the other side [of the 360 degree environment when I looked]. So it's like, I didn't move around as much [afterwards because] I didn't expect there to be something very important." (Smartphone, Greg's perspective, male)*

A few participants suggested that visual directional signals could be added to the immersive story interface to help users identify cue sources faster and therefore minimize attention redirected from their current focus.

**Engagement in the Behavior.** Greg's engagement in binge drinking was conveyed through frequent transitions to a black screen with a table that progressively had more drinks added to it throughout the story, with time stamps associated with the drinks. Some participants reported confusion with this format of information. This was usually due to doubts as to whether Greg was actually consuming all of these drinks, or merely being offered them by friends and potentially not finishing them.

*"They were trying to symbolize how much he was drinking with a black screen and like drinks plopped down, obviously. That was, I feel like we didn't visually see that when we're seeing it from his perspective. [...] I didn't see him drinking that much. Where I was like, this doesn't seem to be following like, how much he's actually drinking compared to what they're trying to illustrate him drinking. I think it's just for*

*visual factors like, beer! Shot! Is he really drinking all of that?" (Smartphone, Greg's perspective, female)*

Some participants admitted having *"no clue"* as to how much alcohol Greg actually consumed prior to hospitalization, which complicated their understanding of binge drinking. Some also felt the black screen *"cut scenes"* were interruptions that detracted from immersion in the story. A few suggested that removing these cut scenes and instead being able to witness Greg consume the drinks would have been a superior choice for conveying the binge drinking behavior.

*"The cuts in between where they told you what time it was kind of un-immersed me from it a little bit. Okay. I feel like if they didn't do that, and just had me following the one character the whole time without interruption, might have been a little bit more convincing." (Smartphone, male)*

**Health Consequences of the Behavior.** The alcohol poisoning and hospitalization experienced by Greg at the end of the immersive story were met with some skepticism and confusion by participants. This was sometimes attributed to an overly rapid transition from *"drunk to dead"* which did not match participants' personal experiences with excessive drinking, and led some to suspect that factors other than binge drinking contributed to Greg's hospitalization.

*"It didn't really seem totally convincing that he just went from drunk to dead. So maybe that's why I didn't think it was binge drinking because it didn't seem as drastic as we usually picture when you're doing something like that." (Smartphone, Greg's perspective, male)*

*"Because of how fast it was. I'll be very honest. I thought there was like a drug in those drinks because it just took off so quickly. Like I thought maybe he spiked the drinks or something or someone spiked drinks." (VR headset, Greg's perspective, male)*

Relatedly, participants were confused about Greg's alcohol poisoning because of the absence of visible escalating health symptoms. While participants who experienced the story from Greg's perspective routinely noted the blurred and slowed vision as a symptom of his drunkenness, several expected progressively worse symptoms indicative of alcohol poisoning. Vomiting was the most commonly expected symptom.

*"The fact that he was fine, then he got a little blurry and it just went from that to passing out. Like I, that doesn't happen. […] I feel like, when I've witnessed people who drink to excess they get really sick and they like to throw up." (Smartphone, Greg's perspective, female)*

*"I didn't connect with it as much because of the progression. Like alcohol poisoning you usually start throwing up first. Other things happen before you're just like gone."(Smartphone, female)*

Participants from Stephanie's perspective echoed a need for information that Greg had transitioned from drunkenness (which they did not consider in need of intervention) to alcohol poisoning. Some of them wondered aloud if they had missed scenes visualizing symptoms of alcohol poisoning, or if Greg's hospitalization was simply unavoidable given the absence of signs for intervention.

*"I just have a question, [about] Greg's thing [the participant references to the entire room]. Was he showing signs like alcohol poisoning? It's like vomiting and twitching and stuff like that." (VR headset, Stephanie's perspective, female)*

### 4.2 Supporting Character Design

Supporting character design refers to the design of characters who viewers can assume the perspective of, but who do not personally engage in the focal health behavior of the story. In this immersive story's case, Stephanie would be the supporting character. While the immersive story intended to focus participants on Greg's binge drinking, participants who assumed Stephanie's perspective talked more about her behavior during focus groups. They exhibited a tendency to assign Stephanie a "role" in the story, which had implications on how they interpreted her actions and Greg's binge drinking.

The majority of participants understood Stephanie's role to a *"good friend."* This carried an understanding that she would look after Greg and either prevent him from drinking too much, or force him to drink water throughout the party. Almost all of these participants thought that Stephanie failed to effectively take care of Greg; they were more critical of her behavior than Greg's, and in some cases blamed her for Greg's hospitalization.

*"If you're a really good friend or just even a friend in general I just truly believe she would be like, 'hey stop drinking, here's the water' you know. She was good friends with him already. She knew he was drunk, like, a good friend would at least offer him a cup of water, at least check on him to see…" (VR headset, Stephanie's perspective, male)*

The mechanism behind how participants arrived at their understanding of Stephanie's role seemed to have been a comparison to how they would act in similar situations of observing a friend drink to excess. While some used this comparison to admonish Stephanie's lack of intervention, others expressed curiosity about her *"personality"* and wanted to know more about her choices.

*"I would have liked to have known her character better, being put into her shoes, because I still have my whole personality and no idea of her personality. So, I would give my friend water. I would lay next to my friend on the couch. And I don't really understand why she didn't do that." (VR headset, Stephanie's perspective, female)*

A few participants shared in critique of Stephanie, but from an understanding that her character was designed to be the *cause* of Greg's drinking. They all pointed to the lack of attention that Stephanie gave to Greg, which—from their understanding—made him feel lonely and prone to drinking. One participant elaborated on this loneliness by suggesting that Greg was in love with Stephanie.

*"This dude's girl, they're about to fight over what this [relationship] is. It looks like every time he was drinking he was alone and drinking or he just wanted Steph's attention for the most part. That's what it seemed like. And then since he wasn't getting any he just kept drinking. And I'm like this is a sad love story." (Smartphone, Greg's perspective, male)*

Not all participants were critical of Stephanie's actions. Some participants reasoned that being the party host prevented her from giving too much attention to any one party guest. They also made common reference to Greg and Stephanie as *"adults"* who are responsible for their own decisions, and that it would not have been appropriate for Stephanie to try to control Greg's drinking.

*"There's a bunch of people here [at the party] that she can't like, he's a grown man and he doesn't need someone watching him." (Smartphone, Stephanie's perspective, female)*

*"You can't tell your friends what to do because people are going to do whatever they want to do." (Smartphone, Stephanie's perspective, female)*

## 5    Limitations

There are some limitations to this focus group study that should be noted. For one, participants' attitudes towards binge drinking and the immersive story could have been influenced by other participants, meaning the opinions of the most vocal participants—or those first to speak—could have been overemphasized. In addition, while our sample does align with victim demographics of binge drinking [25], the attitudes and perceptions of our sample may not generalize to other viewer demographics, such as those not in college, those from largely minority populations, those outside of the United States, and those outside of the early/mid-20s age range. The study also included only one immersive story, and it is unclear if or how the findings may generalize to other types of immersive stories and for different focal health behaviors. Lastly, it cannot be known with our study's design whether viewers' attitudes towards binge drinking affected their subsequent drinking behavior—we thus reiterate a distinction between attitudes and behavior post-immersive story consumption.

## 6    Discussion

We conducted a focus group study with 147 viewers of an immersive story about binge drinking to identify aspects of immersive story design that influence attitudes towards public health issues, and more broadly to understand how health information is consumed in immersive stories. The study elucidated two aspects of immersive story design critical to viewer attitudes and comprehension of binge drinking: 1) health behavior design, including the context of the behavior, engagement in the behavior, and escalation of health repercussions; and 2) supporting character design: the impact of a character who the viewer can assume the perspective of, but who does not personally engage in the detrimental health behavior, on the viewer's attention and judgment of the behavior.  In this section we reflect on how the findings inform immersive story design practices, and open questions that they raise.

### 6.1    Design Constraints of Story Duration

Immersive stories are touted for their capacity to instill a sense of presence [4] in a virtual environment that enables viewers to develop empathy for virtual characters [17, 30] and experience the unfolding story as if it were personally happening to them [32].

Our participants confirmed this potential and indicated that they *wanted* to feel empathy for the characters, but they often failed to do so. A downside of "being there" in a virtual world is that viewers expect the same depth of information that would be customary in the real world to contextualize and comprehend an experience that is really happening to them. This expectation was most apparent in demands for reasons behind the binge drinking behavior, direct observation of alcohol consumption, and an accurate escalation of alcohol poisoning symptoms—all of which the immersive story failed to provide, according to participants.

An impediment to immersive stories simply adding additional content to satisfy these empathic information needs is video duration. Immersive stories are typically around 10 minutes in duration or shorter. Larsen suggests immersive stories are restricted to this length because of general discomfort with VR headsets [15]. One alternative, also posed by Larsen [15], is a serialized narrative made up of multiple videos, but there is much uncertainty as to whether a viewer would actively select the next video to watch (this can be problematic to viewer comprehension if subsequent videos contain vital health information and context). This problem is likely to persist for viewers on mobile devices as well, who are familiar with shorter-form media on social networking apps and are subject to distraction from their surroundings and phone notifications.

Immersive story designers should consider ways to encapsulate viewer-expected information about a health behavior into a relatively short (linear) duration. This poses particular challenges for conveying public health issues that progress over several hours or even days. Design decisions that are ineffective according to our study are cut scenes that indirectly convey engagement in the focal behavior (e.g., a black screen that quantifies the number of alcoholic beverages consumed) and bypassing typical health symptoms to progress to the penultimate health consequence faster. Designers should, at a minimum, prioritize conveyance of context (i.e., reasons) for the character's behavior, direct engagement in the behavior, and all of its standard escalating health symptoms.

### 6.2 Design Considerations for the Supporting Character Perspective

The immersive story literature has elaborated on a variety of character perspective options that designers can choose from (i.e., the eyes through which a viewer experiences the story). These options include non-character observer, the main character (i.e., the character that engages in the focal behavior), and a supporting character (a character that observes the focal behavior, but does not personally engage in it) [11, 15, 17]. Arguments for the supporting character perspective posit that it enables viewers to develop empathy for others as opposed to oneself (Kors et al., 2018), and facilitates explorability of the virtual environment without losing narrative coherence because the supporting character is not personally engaging in plot-driving acts [15]. These arguments frame the supporting character purely as an observer whose own actions have no consequence on the story or the viewer's comprehension of it. Our findings indicate that supporting characters play a more active role in the eyes of viewers.

In our study's immersive story, the supporting character was Stephanie, a party host who observed the main character's (Greg's) binge drinking behavior and found him unresponsive from alcohol poisoning. Our participants were just as attentive to Stephanie's actions as to Greg's, and in many cases were more critical of her lack of intervention into Greg's drinking than the actual binge drinking behavior. Merely observing a main character's behavior, in this sense, was considered a (non-)action in itself, or a conscious behavioral choice from the supporting character. It became the focal point of viewers' attention when the character's absence of intervention into the binge drinking behavior deviated from how viewers would have acted in that situation. This finding emphasizes that immersive story designers need to be attentive to supporting character design, and recognize that passive observance of a situation can be considered the character's conscious behavioral choice.

The sense of embodiment that viewers may feel when they view a story through a character's eyes has traditionally been seen as an advantage of VR experiences [41–44]. However, a non-character observer perspective that removes the viewer from any sense of embodying a character may be superior if the designer wishes the viewer to passively observe a situation. Otherwise viewers may redirect their attention to how their embodied character does or does not utilize their opportunity to alter the events unfolding.

### 6.3   Designing Attention Cues

Managing viewer attention has been a persistent concern in the immersive story literature [10, 12, 14, 20, 21]. Story designers and viewers fear that vital plot points and information will be missed if the viewer exercises their ability to visually explore their 360° field of vision. Our participants certainly shared in this concern. Designs for directing viewer attention have been posed and tested in prior research [21, 22], typically in regards to how the attention cue is generated (e.g., visual cue, audio cue). Our study introduces another dimension to attentional design: the viewer-perceived value of the information being cued.

The binge drinking immersive story viewed by our participants featured a variety of audio cues, but some participants reported ignoring later cues because the information gathered when they sought out prior cues was considered uninteresting or inapplicable to the plot. This begs the question: should immersive story designs attempt to redirect the viewer's attention at all for non-essential plot information? If viewers are dismayed by the value of early nonessential cues, they may miss out on other, plot-imperative cues later. We therefore argue that immersive story designers should minimize non-essential attention cues, and consider how to incorporate plot-essential cues in ways that are not perceived to conflict in priority with whatever is in the viewer's current field of vision.

### 6.4   Future Work

The aforementioned discussion points give researchers and designers several avenues for exploration. Future work should involve viewers in assessment of variations to supporting character design and health behavior design, as well as continued identification of latent or overlooked immersive story elements influential to viewer perception. Considering that the present study focused only on viewers' attitudes towards binge drinking, and immediately after experiencing the immersive story, future work can and should explore long term impacts of immersive story experience on viewers' actual behaviors regarding the focal health issue.

# 7 Conclusion

Immersive stories for health—or 360° videos that convey narrative about a public health issue— have the potential to influence public health at scale because they enable viewers to viscerally experience the consequences of a detrimental health behavior without experiencing real harm. This paper reported a focus group study that facilitated viewers of an immersive story about binge drinking in elucidating elements of immersive story design most impactful to their attitudes towards the respective health behavior. Findings provide insight on: 1) health behavior design: how context of the behavior is sought by viewers, how engagement in the behavior is perceived, and how health repercussions are comprehended; and 2) supporting character design: the impact of a secondary character's behavior on the viewer's attention and judgment of the focal health behavior. Ultimately, while immersive stories hold promise for positively impacting public health, the study demonstrates that immersive stories can have minimal or adverse impact on viewer attitudes if their design is not closely considered.

## Acknowledgements

We thank Devin Yang, Stephen Davidson, and Rukkmini Goli for their data analysis contributions. We also thank Ryan Handley for his artistic contributions in creating Figure 1 for this paper.